\documentclass[10pt]{article}
\usepackage{graphicx}
\usepackage{amsmath}
\usepackage{amsfonts}
\usepackage{amssymb}
\usepackage{cite}
\usepackage{achemso}

\begin{document}
\title{Semi-flexible chain condensation by neutral depleting agents: role of correlations between depletants}
\author{M. Castelnovo\thanks{present address: Laboratoire de Dynamique
des Fluides Complexes,
3 Rue de l'Universit\'e, 67000 Strasbourg, FRANCE,
Email:martin.castelnovo@ldfc.u-strasbg.fr} and W. M. Gelbart \\
\textit{Department of Chemistry and
Biochemistry},\\ 
\textit{University of California Los Angeles,}\\ \textit{Los Angeles, California
90095}}
\maketitle
\begin{abstract}
We revisit the problem of semi-flexible chain condensation by neutral depleting agents (\textit{e.g.} colloidal spheres or flexible polymers) by using a simple formalism that allows us to address its main features without specifying the kind of depleting agents. Correlations between depleting agents are shown to produce a reswelling of the chain at high enough volume fraction, consistent with an earlier analysis by Grosberg \textit{et al.} (\textit{Biopolymers}, 21:2413, 1982) in the context of DNA condensation induced by neutral polymers. It is also shown that the conditions under which spherical colloids can condense a semi-flexible chain are strongly restrictive, unlike what has been recently claimed. The formalism allows us to compare the efficiency of colloids \textit{vs} polymers as condensing agents.
\end{abstract}
\section{Introduction}
Since the early and pioneering work of Asakura and Oosawa \cite{asakura} in colloid-polymer mixtures, depletion forces have been invoked to explain a number of behaviors in various systems. In the original system considered by Asakura and Oosawa, for example, or in the case of an asymmetric binary mixture of hard spheres, the effective interaction between pairs of large spheres is attractive at short distances due to the depletion of the smaller component in between and can induce fluid-fluid phase separation \cite{bibette,poon,dijkstra}. The theoretical description of depletion potentials has even recently become still more quantitative for the simplest geometries like the asymmetric hard sphere mixture \cite{evansgen}. In particular, it has been realized that accurate depletion potentials must take into account correlations among depleting agents.  Those correlations give rise to an energy barrier in the effective interaction between pairs of large colloids that they have to cross before feeling the classical attractive (Asakura-Oosawa) potential. This feature strongly affects the phase diagram of asymmetric mixtures \cite{review_depl}.

Depletion forces are also responsible for the \textit{polymer salt induced} (psi) or $\psi$ condensation at the single chain level: each isolated chain of semi-flexible polyelectrolyte is condensed into a dense toroid above some critical concentration of neutral polymer and salt. Since its discovery in the seventies by Lerman, in the particular case of DNA and polyethyleneglycol (PEG), this condensation has been addressed extensively both experimentally and theoretically 
\cite{lerman,laemmli,maniatis,igor,grosberg,bloomfield1991,khokhlovcollapse,odijkpsi,odijkshape,khokhlov,devries}. Its physical origin is attributed to depletion interactions between DNA segments induced by PEG; the volume occupied by the DNA depletes a certain amount of PEG molecules and therefore induces an effective attraction between DNA segments. The theoretical methods used so far in the context of single chain condensation are usually taken from polymer physics theories. Due to the connectivity of chain segments, correlation effects are known to be very important both at the single chain and solution level \cite{degennes}. Consequently, the emphasis has mainly been put so far on these correlations when describing the chain collapse \cite{khokhlovcollapse,grosberg,khokhlov,devries,searcollapse,vds,odijknucleoid1}. In particular, it has recently been suggested that spherical colloids of appropriate sizes could also condense DNA \cite{searcollapse,vds}. In this spirit, Odijk proposed a model for the nucleoid of bacterial cells \cite{odijknucleoid1,odijknucleoid2}, wherein cellular DNA is condensed by the depletion interactions between DNA segments induced by proteins of the cytoplasm that bear the same charge as DNA. In all these works, the correlations between colloids were limited to the second virial level; also, no restrictions on the colloid size were necessary to obtain the DNA condensation. However, as first emphazised by Grosberg \textit{et al.}, the omitted correlations between PEGs in the case of $\psi$-DNA condensation can qualitatively change this picture \cite{igor} -- above another, higher, critical concentration of PEG the DNA condensate reswells. One speaks then of a 'reentrant condensation'. This phenomenon has indeed been observed by Vasilevskaya \textit{et al.} by fluorescence microscopy \cite{khokhlovcollapse}, and is very similar to what is predicted and observed for gels immersed in polymeric solution where there is a collapse and reswelling of the gel for two different critical polymer concentrations in the solution \cite{khokhlovgel}. 

Apart from the work of Grosberg \textit{et al.}, the level of description of depletion effects in single chain condensation contrasts with that in the recent literature on the phase behavior of colloid-polymer mixtures, which latter is mainly described within the formalism of simple liquid theories where the correlations between depleting agents are highlighted \cite{evansgen,review_depl}. As will be shown in the present paper, insufficient account of correlations in the single chain collapse problem can lead to missing important effects. In particular, inclusion of correlation effects between depleting agents in the case of semi-flexible chain condensation by neutral polymers leads to the requirement of a minimal degree of polymerisation for the polymer to be able to condense the chain, and for re-entrance \cite{igor}. This prediction has been checked experimentally. In the case where spherical colloids are the depleting agents, as discussed below, the effective interactions between chain segments might not become attractive at all, whatever the concentration of colloids used.

The aim of this work is to revisit the effective interactions between chain segments induced by depleting agents, by mixing polymer physics tools and simple liquid theories. We propose a simple formalism that accounts for the main features of semi-flexible chain condensation and treats at the same time a high level of correlations between depleting agents. In the next section, we present the basic framework of our model, without specifying the kind of depleting agents used. Then the application of this model and some quantitative results for hard spheres are given in the third section. Finally, those results are discussed and compared to previous approaches to semi-flexible chain collapse, and in particular DNA condensation. Moreover, the relative efficiency of flexible polymer \textit{vs} colloid for chain condensation is compared.

\section{General results}

\subsection{General framework for the description of semi-flexible chain condensation}
The semi-flexible chain will be modeled as a freely jointed chain, of segment length $l_P$ and width $D$ \cite{degennes}. Each segment can be divided into $n$ ``monomers'' of size $D$ and the total number of monomers of the chain is $N$, considered to occupy a spherical volume $V$. We neglect any kind of intramolecular orientational ordering due to the anisotropic shape of the segments, an approximation valid in the limit of infinite chain length as long as the density of the condensate does not exceed the critical density of the isotropic-nematic ordering transition of unconnected segments \cite{semenov}. Since our goal in this work is not to draw a precise phase diagram of chain condensation, but rather to study the effect of correlations on the induced depletion interactions, the choice of an isotropic condensate with an imposed spherical geometry seems well suited for the tractability of the model. In particular, one should be able this way to address the geometric properties of depleting agents (degree of polymerization or colloid size) required to achieve condensation in a finite range of depletant concentration.  

The partition of the neutral depleting agents inside and outside this volume, and their interactions with chain segments, can give rise to effective attractive interactions. The volume fraction of depleting agents inside and outside the chain volume are respectively $y_{in}$ and $y$. To be more specific with the notation, we consider explicitly the case of spherical colloids as condensing agents, but the results that we derive in this section will be easily generalized to other depleting agents like flexible polymers. The direct interaction (per unit volume) between chain segments and colloids is chosen to have the form:
\begin{equation}
\label{coupling} \frac{F_{int}}{V}\equiv\tilde{F}_{int}=\frac{v_{cross}c_{S}y_{in}}{\pi a^3 /6}
\end{equation} 
where $v_{cross}=\frac{\pi}{4}l_{p}(D+a)^2$ is the second cross virial coefficient between chain segment and depleting agents, $c_{S}$ is the segment density inside the chain volume and $a$ is the diameter of the colloids. This choice will be justified when we discuss the results of our model in section \ref{discussion}. All energies are measured in units of $kT=1$. Let us introduce the reduced concentration $X=v_{cross}c_{S}$. The grand potential describing the equilibrium properties of the system is
\begin{equation}
\label{omega} \Omega [V,y_{in}]=V\left\{\tilde{F}(y_{in})+\tilde{F}_{int}(y_{in})+p(y)-\frac{\mu(y)y_{in}}{\pi a^3 /6}\right\} +F_{pol}(X)
\end{equation}
The quantities $\tilde{F}(y),p(y)$ and $\mu (y)$ are respectively the free energy density, osmotic pressure and chemical potential of the pure colloid system. We do not specify right now the form of these quantities, but the Scaled Particle Theory (SPT) for hard spheres \cite{howard1} will be used to evaluate them in the next section. Similarly we do not specify yet the free energy, $F_{pol}(X)$, associated with the pure semi-flexible chain. It will consist of excluded volume effects, and conformational entropy (see Eq.\ref{fpol}). 

The system described by the grand potential $\Omega$ is in equilibrium with a reservoir of colloids and a reservoir of volume, which means that the colloid chemical potential and the pressure of the system are imposed. The equilibrium equations for the ``inside'' and ``outside'' condensing agents are the equality of their chemical potential and the equality of osmotic pressure:
\begin{eqnarray}
\label{mu1}\mu(y_{in})+X & = & \mu(y) \\
p(y_{in})-p(y)+X\frac{y_{in}}{\pi a^3 /6}+P_{pol}(X) & = & 0
\end{eqnarray}
with the osmotic pressure of the condensing agents given by $p(y)=-\tilde{F}(y)+y\frac{\mu(y)}{\pi a^3 /6}$. The pressure due to the semi-flexible chain alone is denoted $P_{pol}$. By replacing $X$ in the osmotic pressure balance with its expression from the chemical potential equilibrium, the osmotic balance condition can be rewritten
\begin{equation}
\label{bal}P_{eff}(X)+P_{pol}(X)=0
\end{equation}
In the last equation, the effective pressure or interaction $P_{eff}$ is defined by 
\begin{eqnarray}
\label{peff} P_{eff}(X) & \hspace{-.2cm}= & \hspace{-.2cm}\tilde{F}(y)-\tilde{F}(y_{in})+\frac{\mu(y)}{\pi a^3 /6}(y_{in}-y)\\
\label{chemicalpot} y_{in} & \hspace{-.2cm}= & \hspace{-.2cm}\mu^{-1}\left[\mu(y)-X\right]
\end{eqnarray}
If the functional form of the condensing agent chemical potential is known, the effective interaction can be estimated by using Eqs. \ref{bal},\ref{peff} and \ref{chemicalpot}. Indeed $\mu(y)$ is the \textit{only} input required to make this estimate. 

In order to obtain very general information on the effective interactions between chain segments, we can compute their properties as a virial expansion in $X$ \cite{igor}. The $n$-th coefficient, \textit{i.e.} the contribution in $X^n$ in the osmotic balance, is obtained by taking the $n$-th derivative of the effective pressure $P_{eff}(X)$, evaluated at $X=0$. The $n=0$ and $n=1$ derivatives vanish identically when evaluated at $X=0$. But the second derivative of $P_{eff}$, when added to that from the ``bare'' term $P_{pol}$, and evaluated at $X=0$, gives the effective second virial coefficient between chain segments
\begin{equation}
\label{veff} v_{Seff}=\frac{1}{2}\, \left(\left(\frac{\partial^2 P_{pol}}{\partial X^2}\right)_{X=0}-\frac{6}{\pi a^3 \mu'(y)}\right)
\end{equation}

With this general form for the effective virial coefficient that depends only on the chemical potential of the pure condensing agent system, we can study the role of correlations between colloids. If the colloid chemical potential is represented as a virial expansion as a first step, for example, one has for hard sphere depleting agents
\begin{eqnarray}
\mu  & \simeq & \log y+8y+15y^2 \nonumber\\
\label{muprime} \mu' & \simeq & \frac{1}{y}+8+30y
\end{eqnarray}
where the exact second and third virial coefficients for hard spheres have been used \cite{howard1}. Now if we stop the expansion at the second virial level for the condensing agents (\textit{i.e.}, keeping only the first two terms in Eq. \ref{muprime} for $\mu'$), the effective second virial of chain segments is a monotonic decreasing function of the colloid volume fraction and becomes negative above a critical value of $y$. In the framework of the coil-globule transition theory, a negative second virial coefficient means that the chain is collapsed from a swollen to globular state where the correlation of segment density fluctuations is much smaller than the globule size \cite{lifshitz}. Accordingly there should be an unconditional collapse of the semi-flexible chain by colloids of any size once the critical concentration is reached. However if the third virial coefficient of colloids is taken into account, and more generally higher order coefficients, the effective second virial coefficient between chain segments is a non-monotonic function of the colloid volume fraction. Therefore the precise value of the minimal effective second virial coefficient determines if there is actually a collapse of the chain, as will be discussed in the next section. 

If this is the case, then the effective second virial is predicted to become positive again for higher volume fraction of colloids: the chain reswells. This behavior can be interpreted in the following way: the interactions between colloids become more and more important as their volume fraction is increased, so that the interactions between chain segments and colloids are only a slight perturbation of the colloidal fluid. Therefore the system tends to come back to the swollen state of the chain, where the colloid volume fraction inside and outside are similar, since colloids do not lose any translational entropy this way. 

Let us illustrate qualitatively these results in the case where flexible polymers are the neutral condensing agents. For concentrations below the overlap concentration for polymers, the chemical potential is simply $\mu \sim \log \frac{c}{M}$, where $c$ is the monomer concentration and $M$ is the molecular weight of the polymer. For semi-dilute concentrations, the chemical potential is $\mu \sim Mc^{5/4}$, taking into account scaling laws of neutral polymers \cite{degennes}. The derivative of the chemical potential therefore reads
\begin{eqnarray}
 & & \mu' \sim \frac{M}{c} \hspace{1.6cm} \,  \mathrm{if } \,  c<<c^{*} \nonumber\\
 & & \mu' \sim M^2 c^{1/4} \hspace{1cm}\mathrm{if } \, c>>c^{*}
\end{eqnarray}
The effective second virial coefficient is non-monotonic as in the case of strongly correlated colloids. We expect also a reentrant collapse of the chain at high concentration of polymers. This reentrant collapse for a single semi-flexible chain interacting with a solution of flexible polymers was predicted for the first time by Grosberg \textit{et al.} \cite{igor} in the context of $\psi$-DNA condensation and, as mentioned in the introductory section, this reswelling of the DNA has been observed experimentally by  Vasilevskaya \textit{et al.} \cite{khokhlovcollapse}. Note that according to these authors, such a reentrant collapse could be taken into account through a concentration dependent Flory parameter for the flexible polymer, which is equivalent to including a higher level of correlations in the system (see below). 

Related results concerning the role of correlations on the effective second virial coefficient in mixtures of hard spheres and also colloid-polymer mixtures have been obtained previously. As mentioned earlier, one of the early models of the depletion potential in colloid-polymer mixtures is due to Asakura, Oosawa and then Vrij \cite{asakura,vrij} . But the interactions between depleting agents (here the polymers) are not taken into account within those models, and the resulting depletion potential between colloids is purely attractive. On the other hand, more accurate treatments of this system, including interactions between depleting agents, give rise to oscillatory potentials with both attractive and repulsive parts \cite{roth1}. In particular, there is a growing repulsive energy barrier to the short range Asakura-Oosawa type attraction as the concentration of depleting agents is increased. Consequently the effective second virial coefficient between colloids is non-monotonic \cite{roth1}, a feature that is also predicted within our simple model.

\subsection{Explicit model for the semi-flexible chain}
The free energy associated with the pure semi-flexible chain system consists of an excluded volume term, and a conformational entropy contribution:
\begin{equation}
\label{fpol}F_{pol}(X)=F_{ev}+F_{conf}
\end{equation}
with
\begin{eqnarray}
F_{ev} & = & \left(\frac{v_{S}}{2v_{cross}^2}X^2 \right) V \nonumber\\
F_{conf} & = & \frac{3}{2}\left(\frac{R^2}{l_p ^2 N/n}+\frac{l_p ^2 N/n}{R^2}\right)\label{barepol}\nonumber
\end{eqnarray}
The excluded volume term is limited for the sake of simplicity to the
second virial term, although full self-consistency within the
approximations made here would require, for example, developing a scaling form
for the unperturbed polymer free energy. But Eq. \ref{barepol}
suffices, as our aim is to concentrate on the role of the effective
second virial coefficient in driving the collapse and swelling transitions. Since the chain segment is approximated by a rod of width $D$ and of length $l_P$, the virial coefficient is given by $v_{S} =\frac{\pi}{2}l_p ^2 D$. The conformational entropy is composed of the cost for stretching (first term) and compressing (second term) an ideal chain.

To determine the equilibrium conditions for the chain in a solution of depleting agent (volume fraction $y$), we need to minimize the grand potential $\Omega$ -- given by Eq.\ref{omega} -- with respect to: \textit{(i)} the number of colloids ``inside'' the chain volume; and \textit{(ii)} the volume occupied by the chain. The first minimization leads, as before, to Eq.\ref{mu1} or -- equivalently -- \ref{chemicalpot}. The second gives again Eq.\ref{bal} but now, with Eq.\ref{fpol} for $F_{pol}(X)$ allowing us to evaluate explicitly $P_{pol}\equiv -(\partial F_{pol}/\partial V)$, keeping all numbers of particles constant. We find
\begin{eqnarray}
& & \hspace*{-1cm}P_{eff}(X)+\frac{v_{S}}{2v_{cross}^2}X^2 - \frac{(3/4\pi)^{2/3}}{(N/n)^{4/3}l_p ^2 v_{cross}^{1/3}}X^{1/3}+ \frac{(4\pi /3)^{2/3}}{(N/n)^{2/3}l_p^{-2} v_{cross}^{5/3}}X^{5/3}=0 \nonumber\\
& & \label{osm}
\end{eqnarray}
The terms adding to the colloid-induced contributions $P_{eff}(X)$ are: the excluded volume associated with the second-virial interaction between chain segments (each with volume $v_{S}$ and number density $c_{S}=X/v_{cross}$); and the chain conformational terms going as $\partial (R^2)/\partial V\sim V^{-1/3} \sim X^{1/3}$ and $\partial (R^{-2})/\partial V \sim V^{-5/3} \sim X^{5/3}$. As emphasized in the previous section, the induced depletion interactions are given by $P_{eff}=p(y_{in})-p(y)+\frac{Xy_{in}}{\pi a^3/6}$ in the osmotic balance, with $y_{in}(X)$ determined from Eq.\ref{chemicalpot}.

Within a virial expansion, the general form of the osmotic balance reads
\begin{equation}
-AX^{1/3}+BX^{5/3}+(C-C_{coll})X^2+D_{coll}X^{3}+...=0
\end{equation}
with ``\textit{coll}'' denoting the contribution from colloid-induced depletion effects. The solutions of this equation are extrema of the grand potential Eq.\ref{omega}. One has therefore to be careful with the stability of the solutions of interest. This form for the osmotic balance can be used to discuss qualitatively the various features of chain condensation and to guide the numerical analysis. The first step of the numerical analysis is to invert numerically the function ``chemical potential'' -- see Eq.\ref{chemicalpot}, so that $y_{in}(X)$ can be used directly in the effective interaction Eq.\ref{osm}, without any virial expansion.

If there are no colloids, the osmotic balance reads $-AX^{1/3}+BX^{5/3}+CX^2=0$. Under good solvent conditions for the semi-flexible chain, the term in $A$ balances the term in $C$, while in a $\theta$-solvent, the term in $A$ balances the term in $B$. One can easily check that this equation is just the Flory model for the end-to-end radius of the chain \cite{degennes}. In the presence of colloids, the depletion interactions induced by the hard spheres change the effective interactions among segments.

\section{Quantitative results for hard sphere-like depleting agents}
\label{numerical}
In this section, we consider explicitly the case of hard spheres as depleting agents. The only input needed to compute the depletion interaction $P_{eff}$ between chain segments in the previous general model are the properties of the pure hard sphere system through the chemical potential. Since we are dealing with a hard sphere fluid, we can use the thermodynamic quantities derived from the Scaled Particle Theory (SPT) which are known to be accurate as long as the colloid volume fraction is not too close to the close packing limit \cite{howard1}. The free energy density, osmotic pressure and chemical potential within the SPT read
\begin{eqnarray}
F(y) & = & \frac{y}{\pi a^3/6}\left(\frac{3}{2(1-y)^2}+\log \frac{y}{1-y}
\right)\\
p(y) & = & \frac{6}{\pi a^3}\frac{y+y^2+y^3}{(1-y)^3}\\
\mu(y) & = & \log \frac{y}{1-y}+\frac{5-y+2y^2}{2(1-y)^3}
\end{eqnarray}
It is well known from hard sphere theory that the second and third virial computed through the SPT are exact, while up to the fifth order the differences are small (less than 5\%). 

\begin{figure}
[h!]
\begin{center}
\includegraphics*[scale=0.75]{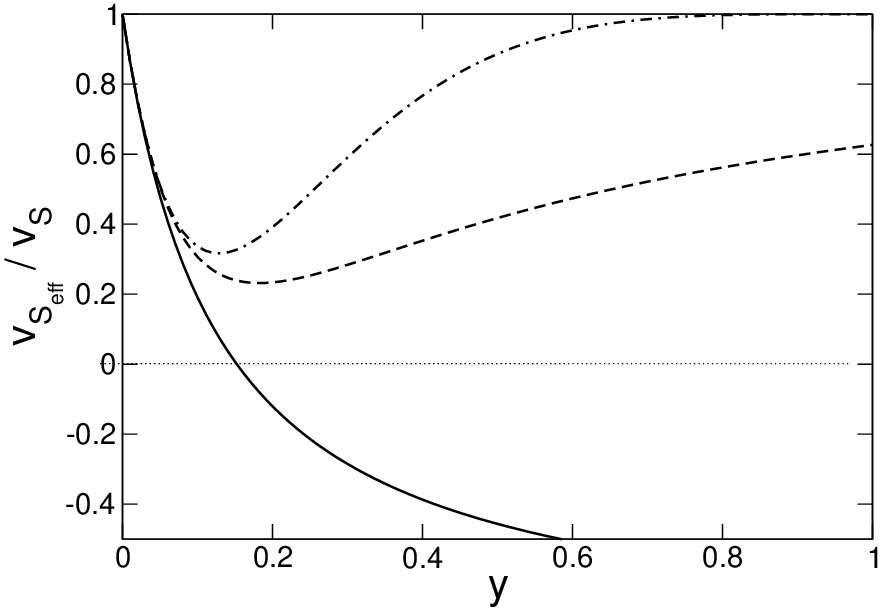}
\caption{Effective second virial coefficient as function of $y$. From bottom to top: second virial approximation, third virial approximation and SPT approximation for the colloids. $a/D=15$}
\label{figveff}
\end{center}
\end{figure}

Using the analysis of the previous section, the nature (repulsive or attractive) of the effective interaction is determined through the effective second virial coefficient between chain segments. The result, calculated analytically according to Eq.\ref{veff}, is shown for example in figure \ref{figveff} for the particular value $a/D=15$ chosen by Sear to study DNA condensation by spherical colloids \cite{searcollapse}. 

As mentioned before, if the pure colloid system is described only up to the second virial level, as in the analysis of Sear, the effective second virial coefficient between chain segments becomes negative above a critical colloid volume fraction ($\simeq 0.17$ in this instance); under these conditions, the chain (DNA) should collapse. However if the pure colloid system is described up to the third order, the minimum of the effective second virial coefficient is positive and there is no collapse of the chain. This result is confirmed when the pure colloid system is described through the SPT, the minimum of the effective second virial coefficient being slightly higher and still positive. Note that the effect of the correlations is relevant for moderate values of colloid volume fraction ($y\approx .2$) so that the description of the hard sphere fluid through the SPT should be valid. 
Also, the volume fraction at which the effective second virial coefficient is minimum does not depend on the ratio $a/D$. It is determined solely by the level of correlations between depleting agents, see, \textit{e.g.} third virial order vs SPT in figure \ref{figveff}.

\begin{figure}
[h!]
\begin{center}
\includegraphics*[scale=0.75]{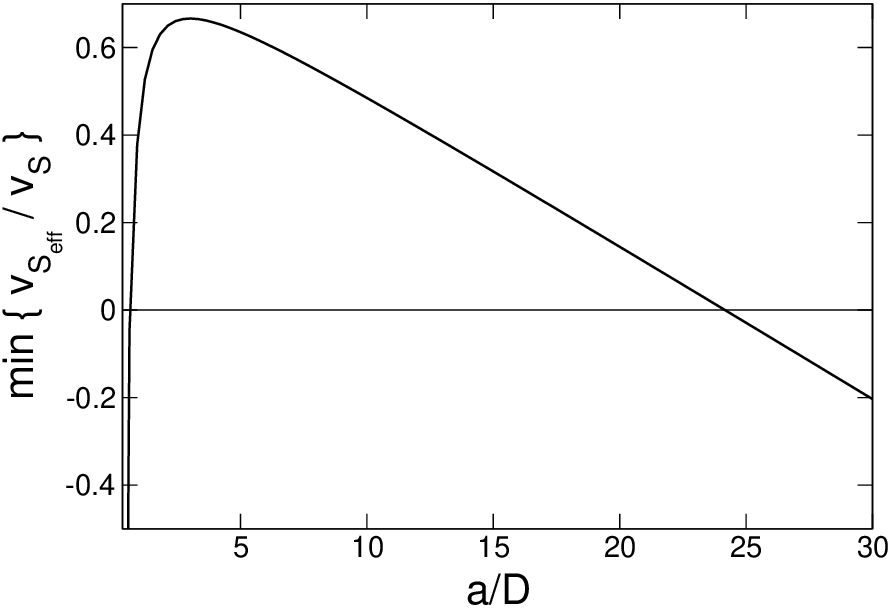}
\caption{Mimimum of the effective second virial coefficient as function of $a/D$.}
\label{figveffmin}
\end{center}
\end{figure}

It is straightforward to compute the minimum of the effective second virial coefficient within the SPT description of the colloids as a function of the ratio $a/D$. The result is shown in figure \ref{figveffmin}. 

We see that the colloids can induce a negative second virial, only if $a/D<0.63$ or $a/D>24.17$. Now for DNA, we have $l_P /D \simeq 25$, so that according to this model colloids whose diameters are bigger than the persistence length of DNA should condense DNA. Unfortunately, we also expect the model to break down in this limit because the form we chose for the cross virial $v_{cross}$ should no longer be appropriate. On the other hand, the range $a/D<0.63$ corresponds to molecular sizes rather than colloids, so that we do not expect any realistic colloids that behave like hard spheres to condense DNA. 
Nevertheless, as is shown immediately below, we still find chain condensation by spherical colloids, but under different conditions. 

\begin{figure}
[h!]
\begin{center}
\includegraphics*[scale=0.75]{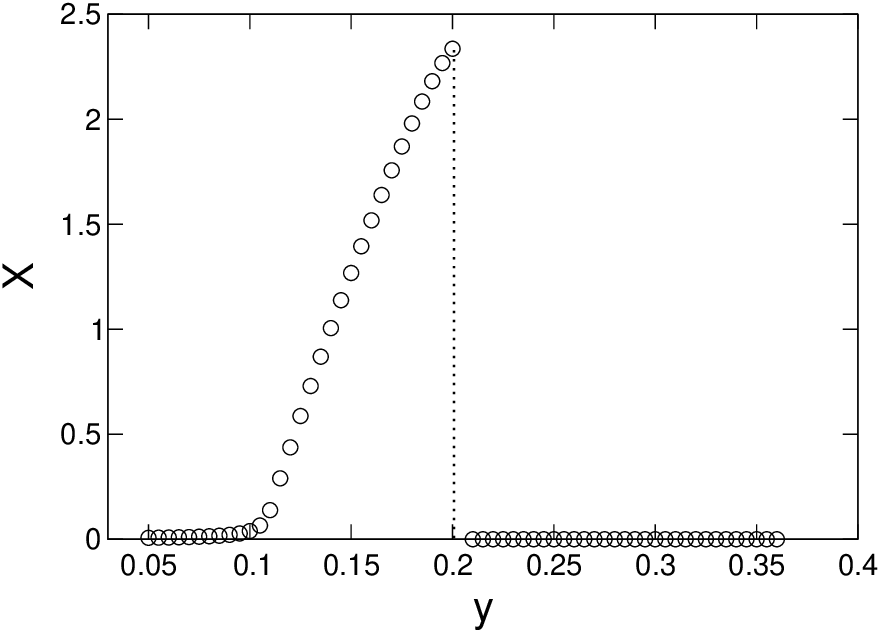}
\caption{Reduced equilibrium density of chain segments $X$ as a function of the colloid volume fraction $y$ in the limit of infinite chain length and reduced solvent quality. $a/D=15\, ,l_p/D=25\, ,\tau=0.67$.}
\label{figxvsy}
\end{center}
\end{figure}

Note that the relative magnitude of the depletion interactions, \textit{i.e.}, the ratio between the induced second virial coefficient and the bare one, scales like
\begin{equation}
\label{magndepl}\frac{v_{cross}^{2}}{v_{S}a^3 \mu '(y)}
\end{equation} 
It is possible to increase the effect of depletion interactions induced by hard spheres by decreasing, say, $v_S$. This corresponds to decreasing the solvent quality for the semi-flexible chain by adding, for example, a certain amount of methanol to aqueous solution \cite{nordmeier1,nordmeier2} in the case of DNA. Of course, this assumes that the colloids still behave purely like hard spheres, so that the solvent does not induce attractive Van der Waals interactions between colloids. Decreasing slightly the quality of the solvent such that the isolated chain is at least in a theta solvent should change the balance between bare and induced interactions. In our model, the solvent quality is taken into account by the substitution $v_S \rightarrow \tau v_S$, where the solvent strength $\tau$ can be varied between 0 (theta solvent) and 1 (athermal solvent). Negative values of $\tau$ correspond to a poor solvent, where the question of chain condensation by depleting agent is irrelevant. The numerical solution of the equilibrium equations in the limit of infinite chain length $(N\rightarrow \infty)$ for $a/D=15$ and $\tau =0.67$, is shown in figure \ref{figxvsy}. It is clear from these results that in this case the chain condensation is continuous, while the reswelling is discontinuous. Note that the density of chain segments in the coil state in the limit of infinite chain length is $X=0$.

\begin{figure}
[h!]
\begin{center}
\includegraphics*[scale=0.75]{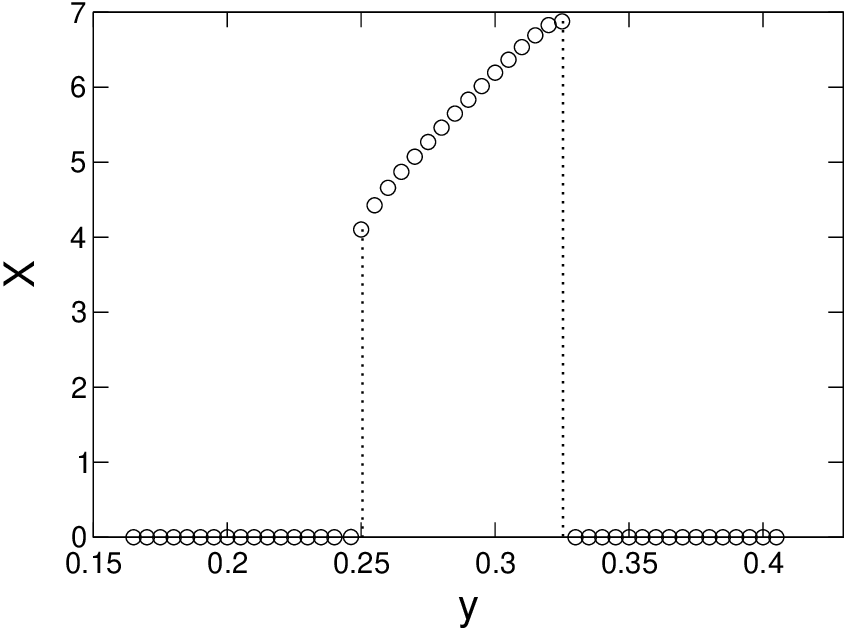}
\caption{Reduced equilibrium density of chain segments $X$ as a function of the colloid volume fraction $y$. $N/n=1.6\,10^{3},\,  a/D=15\, ,l_p/D=25\, ,\tau=0.39$.}
\label{figxvsy2}
\end{center}
\end{figure}

On the contrary, if the chain has a finite length, the conformational entropy term can balance the depletion interactions. In this case, we expect a discontinuous chain condensation; for a given $y$, the repulsive $BX^{5/3}$ term is dominant at low $X$, while the attractive depletion term $-C_{coll}X^2$ is dominant at higher $X$. Above a critical colloid volume fraction, the chain segment density jumps from the coil value to the globule value. As in the infinite chain length limit, the chain condensate reswells to the coil value at still higher volume fraction, because of the strong correlations between colloids. The numerical solution of the equilibrium equations for a particular range of parameters exhibiting this specific behavior is shown on figure \ref{figxvsy2}; accordingly, the solvent quality, \textit{i.e.} $\tau$ has been tuned. Note that the density of the chain segments in the coil state is now finite, although its value is very small (cf figure \ref{figxvsy} \textit{vs} \ref{figxvsy2}).

In both cases of finite or infinite length of semi-flexible chain, we find that the reswelling to the coil state is discontinuous. This can be easily interpreted in terms of effective virial coefficients; above a certain volume fraction of colloids higher than the critical value for chain collapse, the effective second virial coefficient increases with $y$ and becomes positive again while the third virial is still negative. Inspection of higher order induced virial coefficients shows that as the order of the virial coefficient increases, it changes sign more often as a function of colloid volume fraction. This indicates that the use of the virial expansion for the effective interactions is quite limited to the location of a critical value of the colloid volume fraction for chain collapse, and is not well suited to describe precisely the properties of the condensate itself. For example, the reswelling of the chain is not located at the value corresponding to the second change of sign of the effective second virial coefficient. Note that a similar abrupt reswelling is predicted for gels immersed in a solution of polymers \cite{khokhlovgel}. This is not surprising due to the similarities of the starting models describing the same underlying physics of depletion interactions.

\section{Discussion}
\label{discussion}The model we propose in this paper allows us to describe in a simple way both the depletion interactions induced by condensing agents and the effect of correlations between agents on these interactions. Moreover this can be done without specifying the kind of depleting agents. Therefore it can be used to compare the relative efficiency of spherical colloids and flexible polymers for collapsing a single semi-flexible chain. The magnitudes of the induced depletion interactions are respectively
\begin{equation}
\frac{v_{cross}^2}{v_S a^3 \mu '(y)}\mathrm{\ \    and\ \ }\frac{v_{cross}^2}{v_S  \mu '(\frac{c}{M})}
\end{equation}
for colloids and polymers.
In the case of polymers, the value of the cross virial coefficient between a rodlike segment and a polymer coil can be found by requiring a linear dependence in both the length of the segment ($ l_p$) and the degree of polymerization ($M$) \cite{joanny,searrodpol,devries}. The crossed virial term is then given by
\begin{equation}
v_{cross}\simeq l_p D^{1/3}R_g ^{5/3}\simeq l_p D^{1/3}b^{5/3}M
\end{equation}
with $b$ the monomer size of flexible polymer. With this expression one can estimate the strength of the depletion interactions, by using the chemical potential of polymers in a dilute solution:
\begin{equation}
\frac{v_{cross}^2}{v_S  \mu '(\frac{c}{M})}\sim yM\left(\frac{b}{D}\right)^{1/3}
\end{equation}
where $y\sim cb^3$ is the monomer volume fraction. For colloids the equivalent expression reads
\begin{equation}
\frac{v_{cross}^2}{v_S a^3 \mu '(y)}\sim y\frac{\left(1+\frac{D}{a}\right)^4}{D/a}
\end{equation}
where $y$ is the colloid volume fraction. Therefore at equivalent volume fraction there is a factor $M$ favoring the PEG compared to the colloid, and as a rule of thumb flexible polymers should be much more efficient to collapse a semi-flexible chain provided that $M>>1$. More precisely, the polymers are more efficient if
\begin{equation}
M>\frac{\left(1+\frac{D}{a}\right)^4\left(\frac{D}{b}\right)^{1/3}}{D/a}
\end{equation}
The enhanced efficiency of polymers compared to colloids for chain condensation, at comparable volume fraction, is mainly due to: \textit{(i)} the smaller translational entropy compared to the equivalent fluid of monomers and \textit{(ii)} the greater number of interactions between chain segments and monomers.

As mentioned in the introduction, several recent papers have dealt with DNA condensation by depleting agents. In the case of hard sphere depleting agents studied by Sear, the analysis performed in the previous section shows that a more systematic description of correlation effects between colloids rules out simple unconditional DNA collapse above a critical colloid volume fraction \cite{searcollapse}. The conditions required for condensation are strongly restrictive, and within the range of validity of the present theory it happens only for colloids with diameters smaller than that of DNA. Nevertheless, in a real system, this statement can be modified by several factors. First, the polyelectrolyte nature of DNA has not been taken into account in our model. As shown in detail by Ubbink and Odijk \cite{odijkpsi}, this factor strongly influences the shape and density of the condensate. However, since we are mainly interested in the onset of condensation and reswelling, the electrostatic contribution can be included in an effective bare second virial coefficient \cite{joannybarrat}.
Also in the case of prokaryotic cells studied by Odijk \textit{et al.} \cite{odijknucleoid1,odijknucleoid2}, the depleting agents are assumed to be negatively charged proteins, \textit{i.e.}, similarly charged as DNA, and this should increase the value of the cross virial coefficient $v_{cross}$ between DNA segments and proteins, and therefore increase the strength of the depletion interactions according to Eq.\ref{magndepl}. The range of colloid sizes required to condense DNA should be correspondingly larger. The existence of a limited range of volume fractions that can condense DNA should  hold \textit{a priori}; too high a concentration of proteins should not be able to condense DNA, a fact that has not been realized so far. Apart from the depletion effects of those proteins, it is also not clear whether DNA can be considered to be in a good solvent in the cytoplasm \cite{odijknucleoid1,murphy}; this might enhance the depletion interactions between DNA segments induced by the proteins.

The $\psi$-condensation of DNA has also been described recently by comparing the free energy of the system in a coil state and in a condensed state \cite{devries}. This approach neglects completely the nature of the collapse transition and therefore depletion interactions, but still describes qualitatively both the collapse and the reswelling of DNA in PEG solutions. The correlations between PEGs, incorporated through polymer scaling laws in both states of DNA (coil and globule), are responsible for the reswelling, as emphasized in our work; the underlying physics is the same, \textit{i.e.}, a balance between the loss in translational entropy of PEG in the collapsed state of DNA and the depletion effects induced between DNA segments.

The main limitation of our work is the choice of the coupling term Eq.\ref{coupling}, which is the simplest way to couple the behavior of the condensing agents and the semi-flexible chain. In particular we assume a linear dependence on the chain segment density; but once the chain is collapsed, one expects this density to increase, and therefore higher order terms in $X$ might be relevant. 
However these terms are expected to favor reswelling, so that there is no need to include them if one is only interested in understanding the qualitative behavior of the system and locating the onset of condensation and reswelling. Some of these terms --\textit{e.g.} $X^2y$ -- can indeed be thought of as renormalizing the bare second virial of chain segments. On the other hand, if higher orders in depleting agent concentration are included in the coupling term, the depletion effect is enhanced. It is hard to evaluate the precise prefactors of such terms, which influence the degree of depletion enhancement. The effect of the cross virial term $Xy^2$, for example,  can be qualitatively understood by an increase of the effective size of the spheres; a third-order cross virial implying one chain segment and two spheres has a very similar effect on the free energy as a second cross virial between a chain segment and a bigger sphere. More generally we believe that the effect of all higher order coupling terms $X^{\alpha}y^{\beta}$ will not change the qualitative picture drawn from our model. 

This conclusion is further justified by a recent study on mixtures of hard rods and hard spheres based on an integral-equation formalism, which shows very similar results for the effective second virial coefficients between rods \cite{schweizer}; in the case of rods dissolved in a hard sphere fluid, the effective second virial between rods is a non-monotonic function of the hard sphere volume fraction. Moreover, the volume fraction $y_{m}$ where the effective second virial coefficients between chain segments is minimum, does not depend on the the ratio $a/D$, a feature that is provided also by our model even though the two models differ in the precise estimate of the minimum. Within the framework of integral-equation theories, the non-monotonic behavior is explained by the potential of mean force, \textit{i.e.} the depletion potential between two rods that shows a growing repulsive barrier to the short-range attractive well associated with the depletion interactions, as the sphere volume fraction is increased, similar to the case of hard sphere mixtures already mentioned. Moreover, these authors observe, as we do, changes of sign of higher order effective virial coefficients that limit the practical use of such a virial expansion. Since integral-equation theories take into account higher order crossed correlations which we do not, it seems that our choice of the lowest-order coupling term suffices to treat the relevant correlations in the system, namely the correlations between condensing agents. Similarly, earlier integral-equation studies, also by Schweitzer and coworkers, on the depletion interactions between proteins immersed in a polymer solution, exhibit a non-monotonic dependence of the effective protein second virial coefficient as the concentration of polymers is increased \cite{schweizer2,schweizer3}. Our present formalism could be generalized to compute analytically this effective depletion interaction between spherical colloids immersed in a polymer solution. The non-monotonic dependence of the second virial coefficient can then be explained by the correlations in the polymer solution.

In the numerical analysis presented in section \ref{numerical}, we chose also the simplest form for the free energy of the pure semi-flexible chain. This is not really a restrictive assumption, because taking into account more correlations between chain segments should contribute also to an early reswelling. 

\section{Conclusion}
We propose in this paper a simple model for semi-flexible chain
condensation by neutral depleting agents. It is shown that the kind of
agents need not be specified to obtain the main features of the
induced depletion interactions between chain segments. In particular,
this allows us to address the role of condensing agent correlations on
the effective interactions. In the case of polymeric depleting agents
like PEG, this effect has already been taken into account by Grosberg
\textit{et al.} \cite{igor}; the effective interaction between DNA
segments is a non-monotonic function of the monomer volume
fraction. Accordingly, we expect a reswelling of the DNA at high
enough monomer concentration, and a lower bound for the PEG length
required to condense DNA. In the case of spherical colloids, our model
predicts the same feature, unlike recent work where correlations have
been neglected. The effective interactions are non-monotonic as a
function of colloid volume fraction, and this again imposes strongly
restrictive conditions on the size of colloids required to condense a
semi-flexible chain. For colloids interacting only through their hard
cores, no realistic range of colloid size is found. Nevertheless, if
the quality of the solvent (aqueous solution) for the chain is
decreased, the range of colloid size should be experimentally
accessible. We gave in section \ref{numerical} an example of such a
situation by solving the equilibrium equations within an SPT
description of the colloids. Our model can also describe in principle
the example of DNA condensation by negatively charged proteins
considered by Odijk; the only input required is the chemical potential
of the pure protein solution. In this case, one expects the cross
virial term between DNA segments and proteins to be larger than the
pure hard core value considered here, due to electrostatic
interactions; the strength of the depletion interaction and therefore
the condensing range of colloid size are increased. We hope that this
work will motivate further experimental studies to determine whether 
DNA collapses in the presence of small colloids (very likely proteins).

The qualitative results obtained in this article are confirmed by
recent integral equation treatments of similar systems
\cite{schweizer,schweizer2,schweizer3}. The analytical model proposed
in this paper provides a simple alternative description of the effect
of correlations between depleting agents on the interactions mediated
by them. This approach can be seen as complementary to the integral equations theories, since it
allows us, in a simple and direct way, to investigate the influence of solvent
quality and colloid charge, etc., on the phase behave of this system.

\textbf{Acknowledgment}

We would like to acknowledge R. Sear for attracting our attention to the role of higher order coupling terms between the DNA and the depleting agents. This work was partially supported by NSF grant \# CHE9988651 to W.M.G.

\end{document}